\documentstyle[12pt]{article}

\def\nqq{\hspace{-2em}}

\def\barr{\left(\begin{array}}
\def\earr{\end{array}\right)}
\def\beq#1{\begin{equation}\label{#1}}
\def\eeq{\end{equation}}
\def\ber#1{\begin{eqnarray}\label{#1} &&\nqq}%   left alignment
\def\eer{\end{eqnarray}}
\def\eern{\nonumber \end{eqnarray}}%   nonumber
\def\mm{\\ &&\nqq}

\catcode`\@=11 \@addtoreset{equation}{section}\catcode`\@=12

\newcommand{\R}{\mbox{\bf R}}

\newcommand{\btd}{\bigtriangledown}

\newcommand{\eps}{\varepsilon}
\newcommand{\tri}{\triangle}

\newcommand{\e}[1]{\mathop{\rm e}\nolimits^{#1}}

\newcommand{\p}{\partial}

\begin{document}

\thispagestyle{empty}

\begin{center}
\large\bf
On Symmetries of Target Space for Sigma-model
of $p$-brane Origin\\[1.5cm]
\normalsize
V. D. Ivashchuk \\[1cm]
Center for Gravitation and Fundamental Metrology\\
VNIIMS, 3-1 M. Ulyanovoy Str.\\
Moscow, 117313, Russia\\
e-mail: ivas@rgs.phys.msu.su
\end{center}

%\bigskip

\begin{abstract}
The target space ${\cal M}$ for
the sigma-model appearing in theories with p-branes
is considered. It is proved that  ${\cal M}$ is a homogeneous space
 $G/H$. It is  symmetric if and only if
the $U$-vectors governing the sigma-model metric
are either coinciding or mutually orthogonal. For nonzero
noncoinciding $U$-vectors the Killing equations are solved.
Using a block-orthogonal decomposition of the set of the $U$-vectors
it is shown that under rather general assumptions
the algebra of Killing vectors is a direct sum of several copies of
$sl(2,\R)$ algebras (corresponding to 1-vector blocks), several
solvable Lie algebras (corresponding to multivector blocks) and
the Killing algebra of a flat space.
The target space manifold is decomposed in
a product of $\R^m$, several 2-dimensional spaces
of constant curvature (e.g. Lobachevsky space, part of
anti-de Sitter space)
and several solvable Lie group manifolds.
\end{abstract}

\pagebreak

%%%%%%%%%%%%%%%%%%%%%%%%%%%%%%%%%%%%%%%%%%%%%%%%%%%%%%%%%%%%%%%%
\section{Introduction}
%%%%%%%%%%%%%%%%%%%%%%%%%%%%%%%%%%%%%%%%%%%%%%%%%%%%%%%%%%%%%%%%

In this paper we consider the metric
\ber{1.1}
{\cal G}= \hat G_{AB} dx^A\otimes dx^B+
\sum_{s\in S} \eps_s \e{2U_A^s x^A} d\Phi^s\otimes d\Phi^s,
\eer
defined on $\R^K$, where $x = (x^A) \in \R^N$,
$\Phi=(\Phi^s, s \in S) \in \R^{|S|}$ ($S\ne\emptyset$),
$(\hat G_{AB})$ is a symmetric non-degenerate matrix,
$\eps_s= \pm1$, $U^s=(U_A^s) \in \R^N$ are vectors, $s\in S$,
and $K = N + |S|$. The  space ${\cal M}=(\R^K,{\cal G})$
is a target space for the $\sigma$-model appearing in
multidimensional theories with intersecting $p$-brane solutions
\cite{IM,IMR}.

Here we prove that  ${\cal M}$ is a homogeneous space
isomorphic to $G/H$ \cite{helgason},
where $G$ is the isometry group of ${\cal M}$,
and $H$ is an isotropy subgroup of $G$ (Sect. 2). We prove that
${\cal M}$ is a symmetric space, iff (if and only if) any two vectors
$U^{s_1}$ and $U^{s_2}$, $s_1,s_2 \in S$, $s_1 \neq s_2$,
are either coinciding $U^{s_1}=U^{s_2}$ or orthogonal
$(U^{s_1},U^{s_2})=0$, with respect to the scalar product
\ber{1.2}
(U,U')=\hat G^{AB} U_A U'_B,
\eer
where  $U,U'\in\R^N$, $(\hat G^{AB})=(\hat G_{AB})^{-1}$. The Killing
equations are solved in Sect. 3. In Sect. 4 the block-orthogonal
decomposition of the set of $U^s$-vectors is considered and
under rather general assumptions the decomposition  of the
Killing algebra into a sum of mutually commuting subalgebras is
obtained.

We note that recently some new classes of solutions
with $p$-branes were obtained in the block-orthogonal case
\cite{Br,IM2}.

%%%%%%%%%%%%%%%%%%%%%%%%%%%%%%%%%%%%%%%%%%%%%%%%%%%%%%%%%%%%%%%%
\section{Coset structure}
%%%%%%%%%%%%%%%%%%%%%%%%%%%%%%%%%%%%%%%%%%%%%%%%%%%%%%%%%%%%%%%%

The isometry group $G = {\rm Isom} ({\cal M}$) has two Abelian
subgroups defined by transformations
\ber{2.1}
\Phi^s \mapsto \Phi^s + \Delta^s, \qquad x^A \mapsto x^A,  \mm
\label{2.2}
\Phi^s \mapsto \Phi^s \e{- U_A^s \delta^A}, \qquad
x^A \mapsto x^A + \delta^A,
\eer
($\Phi^s$-translation and generalized $x^A$-translation respectively)
$\Delta^s, \delta^A \in \R$.

{\bf Proposition 1.}
The  space ${\cal M}=(\R^K,{\cal G})$ with the metric defined
in (\ref{1.1}) is a homogeneous space isomorphic to
$G/H$, where $G = {\rm Isom} ({\cal M}) $ is the isometry group of ${\cal
M}$, and $H$ is the isotropy subgroup of $G$.

{\bf Proof.} It is sufficient to prove that the action of
the group $G$ on $\R^K$ is transitive \cite{helgason}. Indeed,
any two points $a_1, a_2 \in \R^K$ may be ``connected''
by a certain composition of the isometries (\ref{2.1}) and
(\ref{2.2}), i.e. $a_2= f_2(\delta) (f_1(\Delta)(a_1))$,
where the isometries $f_1(\Delta), f_2(\delta):  \R^K \rightarrow \R^K$
are defined in (\ref{2.1}) and (\ref{2.2}) respectively.

Here $H =H_0 \cong H_a$, where
$H_a \equiv \{g| g \in G, \ g(a) = a \}$ is the isotropy
subgroup of $G$ at $a \in \R^K$.

We note that recently  in a special  case of two vectors
(or two $p$-branes)
a similar result was established in the framework of a
$\sigma$-model representation for a non-block-diagonal
metric ansatz \cite{GR}.

{\bf Proposition 2.}  The space ${\cal M}=(\R^K,{\cal G})$
is symmetric, i.e. $\btd_M[{\cal G}] R_{M_1 M_2 M_3 M_4}[{\cal G}] = 0$,
iff
\ber{2.4}
(U^{s_1},U^{s_2}) (U^{s_1} - U^{s_2}) =0,
\eer
for all $s_1, s_2 \in S$.

{\bf Proof.} Relation
$\btd_M R_{M_1 M_2 M_3 M_4} = 0$
is equivalent to the relation
(\ref{A.4}) from the Appendix A, that is equivalent to (\ref{2.4}).

We note that in our previous papers \cite{IM,IMR} we considered
the Majumdar-Papapetrou type solutions for the $p$-brane
$\sigma$-model in the orthogonal case
$(U^{s_1},U^{s_2})=0$, $s_1 \ne s_2$.

{\bf Proposition 3.} Let $U^s \neq 0$ for all $s \in S$.
Then the space ${\cal M}=(\R^K,{\cal G})$ is a space of
constant curvature, i.e.
\ber{2.4a}
R_{M_1 M_2 M_3 M_4}[{\cal G}] =
k ({\cal G}_{M_1 M_3}{\cal G}_{ M_2 M_4} -
 {\cal G}_{M_1 M_4}{\cal G}_{ M_2 M_3}),
\eer
$k = {\rm const}$, iff $N = 1$
and
\ber{2.4b}
U^s_1 = \pm \sqrt{-k \hat G_{11}}
\eer
$s \in S$, where $k \hat G_{11} < 0$.

Proposition 3 can be readily proved using the relations
>from the Appendix A. It follows from (\ref{2.4b})
that $(U^s,U^s) = - k$, $s \in S$.

%%%%%%%%%%%%%%%%%%%%%%%%%%%%%%%%%%%%%%%%%%%%%%%%%%%%%%%%%%%%%%%
\section{Algebra of Killing vectors}
%%%%%%%%%%%%%%%%%%%%%%%%%%%%%%%%%%%%%%%%%%%%%%%%%%%%%%%%%%%%%%%%

Let ${\bf g} = isom({\cal M })$ be the algebra of Killing
vector fields on $M$,
satisfying the Killing equations
\ber{3.1}
\btd_M v_{N} + \btd_N v_{M}  = 0,
\eer
where $v = v^M \p/ \p X^M \in {\bf g}$, $(X^M) = (x^A, \Phi^s)$,
$v_N = {\cal G}_{NM} v^M$, $\btd_M = \btd_M[{\cal G}]$ is the
covariant derivative corresponding to the metric (\ref{1.1}).
The algebra ${\bf g}$ is isomorphic to the Lie algebra of
the isometry group $G$.

{\bf Proposition 4.}  Let
\ber{3.1a}
U^s \neq 0, \qquad  U^{s_1} \neq U^{s_2},
\eer
for all $s, s_1, s_2 \in S$, $s_1 \neq s_2$.
Then the  Killing equations have the following
solutions
\ber{3.2}
v_{A} = C_A + 2 \sum_{s \in S} f_s U^s_{A} \Phi^s + C_{AB} x^B, \mm
\label{3.3}
v_{s} = f_s +  \eps_{s} \e{2U^s_A x^A}[b_s -
U^{s A} C_A \Phi^s - f_s (U^s, U^s) (\Phi^s)^2],
\eer
where constants $C_A, C_{AB}, b_s, f_s \in \R$ satisfy the
relations
\ber{3.4}
U^{s A} C_{AB} = 0, \qquad  C_{AB} = -  C_{BA}, \mm
\label{3.5}
(U^s, U^{s'}) f_{s'} = 0, \qquad s \neq s', \ ({\rm no \ summation})
\eer
$A,B = 1, \ldots, N$; $s, s' \in S$.

We note that the vector fields defined by relations
(\ref{3.2})-(\ref{3.5}) satisfy the Killing equations for arbitrary
$U^s$-vectors.  The sketch of proof of the Proposition 4 is given in
Appendix B.  Here and in what follows we assume that the restrictions
(\ref{3.1a}) are satisfied.

The Killing vector fields corresponding to isometries (\ref{2.1})
and (\ref{2.2}) respectively are the following
\ber{3.6}
T_s = \p_s, \mm
\label{3.7}
D_A = \p_A - \sum_{s \in S} U^{s}_ A \Phi^s \p_s,
\eer
where $\p_s = \p/ \p \Phi^s$ and $\p_A = \p/ \p x^A$.

Let
\ber{3.8}
F_r = 2 U^{r A} \Phi^r \p_A +
[\eps_r e^{-2U^r} -  (U^{r}, U^r) (\Phi^r)^2] \p_r,
\eer
where $r \in \tri \equiv \{ r' \in S| (U^{r'},U^{s}) =0,
\forall s \in S \setminus \{ r'\} \}$.
Thus, the Killing vector field $F_r$ correspond to the vector $U^r$
orthogonal to all other vectors $U^s$.

Let us consider the coordinates $(x^A) = (x^a, x^{\alpha})$
satisfying the relations
\ber{3.10}
\hat G_{AB} = \eta_{AB}= \eta_{AA} \delta_{AB}, \qquad \eta_{AA} = \pm 1, \mm
\label{3.10a}
U^{s}_{\alpha} = U^{s \alpha} = 0,
\eer
$s \in S$; $a = 1, \ldots, p$; $\alpha = p+1, \ldots, N$.

The generators of generalized (pseudo) rotations are
\ber{3.11}
M_{\alpha \beta} =
x_{\alpha} \p_{\beta} - x_{\beta} \p_{\alpha},
\eer
$x_{\alpha} = \eta_{\alpha \alpha} x^{\alpha}$,
$\alpha, \beta = p+1, \ldots, N$.

Commutation relations read
\ber{3.12}
[D_A, D_B] = [T_s, T_{s'}] = [F_r, F_{r'}] = 0, \mm
\label{3.13}
[D_A, F_r] = - U^r_A F_r \ ({\rm no \ summation}), \qquad
[T_s, F_r] = 2 \delta_{sr} U^{r A} D_A, \mm
\label{3.14}
[D_A, T_s] =  U^s_A T_s  \ ({\rm no \ summation}), \mm
\label{3.15}
[M_{\alpha \beta}, M_{\alpha' \beta'}] =
\eta_{\beta \alpha'}  M_{\alpha \beta'}
+ \eta_{\alpha \beta'}  M_{\beta \alpha'}
- \eta_{\alpha \alpha'}  M_{\beta \beta'}
- \eta_{\beta \beta'}  M_{\alpha \alpha'},  \mm
\label{3.16}
[M_{\alpha \beta}, T_s] = [M_{\alpha \beta}, F_r] = 0,  \mm
\label{3.17}
[M_{\alpha \beta}, D_A] =
\eta_{ \beta A}  D_{ \alpha}  - \eta_{ \alpha A}  D_{ \beta},
\eer
$a, b =  1, \ldots, p $; $\alpha, \beta= p+1, \ldots, N$;
$s,s' \in S$; $r \in \tri$.

Let us suppose that the coordinates satisfying
(\ref{3.10}) and (\ref{3.10a}) also obey the relations
\ber{3.18}
p = {\rm rank} (U^s, s \in S).
\eer
In the Euclidean case, when $\eta_{AB}= \delta_{AB}$, such coordinates
do exist.

It follows from the relations
(\ref{3.4}) , (\ref{3.5}), (\ref{3.10}), (\ref{3.10a}) and (\ref{3.18})
that the vector fields
$T_s, D_A, F_r, M_{\alpha \beta}, \alpha < \beta $
($s \in S;  A = 1, \ldots, N; r \in \tri;
\alpha, \beta = p+1, \ldots, N$) form a basis in ${\bf g}$.

%%%%%%%%%%%%%%%%%%%%%%%%%%%%%%%%%%%%%%%%%%%%%%%%%%%%%%%%%%%%%%%%
\section{Block-orthogonal decomposition}
%%%%%%%%%%%%%%%%%%%%%%%%%%%%%%%%%%%%%%%%%%%%%%%%%%%%%%%%%%%%%%%%

Let
\ber{4.3}
S=S_1\sqcup\dots\sqcup S_k,
\eer
$S_i\ne\emptyset$, $i=1,\dots,k$, and
\ber{4.4}
(U^s,U^{s'})=0
\eer
for all $s\in S_i$, $s'\in S_j$, $i\ne j$; $i,j=1,\dots,k$. Relation
(\ref{4.3}) means that the set $S$ is a union of $k$ non-intersecting
(non-empty) subsets $S_1,\dots,S_k$. According to (\ref{4.4}) the set of
vectors $(U^s,s\in S)$ has a block-orthogonal structure with respect to
the scalar product (\ref{1.2}), i.e. it is splitted into $k$ mutually
orthogonal ``blocks'' $(U^s,s\in S_i)$, $i=1,\dots,k$.
We consider the block-orthogonal decomposition to be
irreducible, i.e. for any $i$ the block $(U^s,s\in S_i)$ can not be
splitted into two mutually orthogonal subblocks.

Let us suppose that $(x^a) = (x^{a_1}_1, \ldots , x^{a_k}_k)$,
$a_i = 1, \ldots, p_i$, $i=1,\dots,k$,  and
\ber{4.5}
U^{s_j}_{a_i} = 0,
\eer
$s_j \in S_j$, $i \neq j$; $i,j=1,\dots,k$.
We also put
\ber{4.6}
p_i = {\rm rank} (U^s, s \in S_i),
\eer
$i=1,\dots,k$.
In the Euclidean case ($\eta_{AB}= \delta_{AB}$)
there exist coordinates satisfying
(\ref{3.10}), (\ref{3.10a}), (\ref{4.5}) and (\ref{4.6}).

Let  ${\bf g_0} \equiv < D_{\alpha}, M_{\alpha \beta};
\alpha, \beta = p+1, \ldots, N >$,
${\bf g_i} \equiv < T_{s_i}, s_i \in S_i; D_{a_i}, a_i = 1, \ldots, p_i;
F_{r_i}, r_i \in \tri \cap S_i >$,
where $< (\cdot)>$ is a span of $(\cdot)$ over $\R$.
The Killing algebra is a (direct) sum of
mutually commuting subalgebras:
\ber{4.9}
{\bf g} = {\bf g_0} \oplus {\bf g_1} \oplus \ldots \oplus {\bf g_k},
\qquad [{\bf g_{\nu}}, {\bf g_{\mu}}] = {\bf 0},
\eer
$\nu, \mu = 0, \ldots, k$.

For $|S_i| =1$  ${\bf g_i} \cong sl(2,\R)$.
Indeed, in this case $S = \{ s_i \}$ and
${\bf g} = < T = T_{s_i}, D = D_{a_i}, F = F_{s_i} >$.
In the new basis
$h = 2 U^{-1} D$, $e_{+} = \eta U^{-1} D$,
$e_{-} = U^{-1} F$, where $U = U^{s_i}_{a_i}$ and
$\eta = \eta_{a_i a_i}= {\rm sign} (U^{s_i},U^{s_i})$,
we get the familiar relations for $sl(2,\R)$:
$[h,e_{+}] = 2 e_{+}$, $[h,e_{-}] = - 2 e_{-}$,
$[e_{+},e_{-}] = h $.

For $|S_i| >1$  the Lie algebra ${\bf g_i}$ is solvable,
since ${\bf g_i}^{(2)} = [{\bf g_i}^{(1)}, {\bf g_i}^{(1)}] = {\bf 0}$,
where ${\bf g_i}^{(1)} = [{\bf g_i},{\bf g_i}]$.
We note that recently solvable Lie algebras were
studied for supergravity models in numerous papers
(see \cite{torino} and references therein).

The isotropy subalgebra ${\bf h} = \{v \in {\bf g}| v(0) = 0 \}$
has the following form:
\ber{4.10}
{\bf h} = {\bf h_0} \oplus \sum_{r \in \tri} {\bf h_r},
\eer
${\bf h_0} = < M_{\alpha \beta}; \alpha, \beta = p+1, \ldots, N >$
is the isotropy subalgebra of ${\bf g_0}$ and
${\bf h_r} = <F_r - \theta_r D_{a(r)}>$, where
$\theta_r \equiv \eps_r {\rm sign} (U^{r},U^{r})$, $r \in \tri$.
The subalgebra ${\bf h_r}$ is embedded into $sl(2,\R)$
as $so(2, \R)$ for $\theta_r = +1$, and as
 $so(1,1, \R)$, for $\theta_r = -1$ (this follows from the matrix
realization of $e_{+}, e_{-}, h$).

Under assumptions mentioned above the target space
manifold may be decomposed as follows
\ber{4.11}
\R^K = \R^{N-p} \times \prod_{r \in \tri}  M_r
\times \prod_{l \in \tri_{+}} G_l,
\eer
where  $\tri_{+} = \{l =1,\ldots,k| |S_l| > 1 \}$, $G_l =
\exp({\bf g_l})$ is a solvable Lie group, $l \in \tri_{+}$, $M_r$ is
2-dimensional space of constant curvature isomorphic to the Lobachevsky
space $H^2 = SL(2, \R)/SO(2)$ for $\theta_r = +1$ and to the part of
anti-de Sitter space $AdS^2 = SL(2, \R)/SO(1,1)$ for $\theta_r = -1$. The
target space metric is a direct sum of metrics
\ber{4.12}
{\cal G} = {\cal G}_0 \oplus
\sum_{r \in \tri} \eta_r {\cal G}_r \oplus
\sum_{l \in \tri_{+}} {\cal G}_l,
\eer
where ${\cal G}_0 = \eta_{\alpha \beta} dx^{\alpha} \otimes dx^{\beta}$,
${\cal G}_r$ is the canonical metric on $M_r$, $\eta_r =
{\rm sign} (U^{r},U^{r})$,$r \in \tri$,
and ${\cal G}_l$  is a non-degenerate metric on $G_l$, $l \in \tri_{+}$.

\section{Appendix A}

The nonzero components of the Christoffel symbols
for the metric (\ref{1.1}) are the following \cite{IMS}
\ber{A.1}
\Gamma^{s}_{As} = \Gamma^{s}_{sA} = U^s_A, \qquad
\Gamma^{A}_{ss} =  - U^{sA} \eps_s e^{2U^s}.
\eer

The nonzero components of the Riemann tensor corresponding
to (\ref{1.1}) read
\ber{A.2}
R_{A_1 s A_2 s} = {\rm permutations}=
- U^{s}_{A_1} U^{s}_{A_2} \eps_s e^{2U^s},  \mm
\label{A.3}
R_{s_1 s_2 s_3 s_4} =
\eps_{s_1} \eps_{s_2} (U^{s_1}, U^{s_2})
e^{2U^{s_1} + 2U^{s_2}}
(\delta_{s_1 s_4} \delta_{s_2 s_3} -  \delta_{s_1 s_3} \delta_{s_2 s_4}).
\eer
Here $U^s = U^s_A x^A$ and the scalar product
$(\cdot, \cdot)$ is defined in (\ref{1.2}).

It follows from (\ref{A.1})-(\ref{A.3}) that
the only nontrivial components of the relation \\
$\btd_M R_{M_1 M_2 M_3 M_4} = 0$
(up to permutations of indices) are the following
\ber{A.4}
\btd_s R_{s_1 s_2 s_3 A} =
\eps_{s_1} \eps_{s_2} (U^{s_1}, U^{s_2})
e^{2U^{s_1} + 2U^{s_2}} (U^{s_2}_A - U^{s_1}_A)
(\delta_{s s_1} \delta_{s_2 s_3} + \delta_{s s_2} \delta_{s_1 s_3})
= 0.
\eer

\section{Appendix B}

The Killing equations (\ref{3.1}) read
\ber{B.1}
\p_A v_B + \p_B v_A = 0, \mm
\label{B.2}
\p_A v_s +  \p_s v_A  - 2 U^{s}_A v_s = 0, \ ({\rm no \ summation}) \mm
\label{B.3}
\p_s v_{s'} + \p_{s'} v_s = 0,  \quad s \neq s',  \mm
\label{B.4}
\p_s v_{s} + \eps_s e^{2U^s} U^s_A v^A = 0.
\eer

The first equation has the solution
\ber{B.5}
v_{A} = C_A (\Phi) + C_{AB}(\Phi) x^B,
\eer
where $C_{AB}(\Phi) = - C_{BA}(\Phi)$, $A, B =1, \ldots, N$.
The Proposition 4 may be readily proved using the relations
(\ref{B.2})-(\ref{B.4}), identities
$\p_M \p_N (v_s \e{-2U^s_A x^A}) = \p_N \p_M (v_s \e{-2U^s_A x^A})$
and the following Lemmas.

{\bf Lemma 1.}  Let $(U_A) \in \R^N$, $(U_A) \neq 0$ and
\ber{B.6}
D  = e^{U_A x^A} (B + C_A x^A)
\eer
for all $x \in \R^N$. Then  $B = D = C_A = 0$
for $A = 1, \ldots, N$.

{\bf Lemma 2.}  Let $(U_A) \in \R^N$,
$(U_A) \neq 0$, $C_{AB}= - C_{BA}$, $A,B = 1, \ldots, N$,
$f = f(x)$ and
\ber{B.7}
\p_A f  = e^{U_B x^B} (B_A + C_{AB} x^B)
\eer
for all $x \in \R^N$. Then $C_{AB} = 0$ and $B_A = \lambda
U_A$ for some $\lambda \in \R$.

\begin{center}
{\bf Acknowledgments}
\end{center}

This work was supported in part
by the DFG grant  436 RUS 113/236/O(R),
by the Russian Ministry of
Science and Technology and  Russian Foundation for Basic Research,
grant 98-02-16414.
The author is grateful  to organizers of
Cosmion meeting (December 7--12, 1997) where the results of this
work were reported and to V.N. Melnikov and M.I. Kalinin for useful
discussions.


\small
\begin{thebibliography}{99}

\bibitem{IM}
V.D. Ivashchuk and V.N. Melnikov,
Intersecting p-brane Solutions in Multidimensional
Gravity and M-theory, hep-th/9612089;
{\it Gravitation and Cosmology} {\bf 2}, No 4, 204 (1996);\\
V.D. Ivashchuk and V.N. Melnikov,
{\it Phys. Lett. } {\bf B 403}, 23 (1997); \\
V.D. Ivashchuk and V.N. Melnikov,
Sigma-model for the Generalized  Composite p-branes,
hep-th/9705036; {\it Class. Quantum Grav.} {\bf 14} (11), 3001 (1997).

\bibitem{IMR}
V.D. Ivashchuk, V.N. Melnikov and M. Rainer,
Multidimensional $\sigma$-models with Composite Electric p-branes,
gr-qc/9705005; {\it Gravitation and Cosmology} {\bf 4}, No 2 (14) (1998).

\bibitem{helgason}
S. Helgason, Differential Geometry and Symmetric Spaces,
New York: Academic Press (1962).

\bibitem{Br}
K.A. Bronnikov, Block-orthogonal Brane systems, Black
Holes and Wormholes, hep-th/9710207.

\bibitem{IM2}
V.D. Ivashchuk and V.N. Melnikov,
Mujumdar-Papapetrou Type Solutions in  Sigma-Model
and Intersecting p-Branes, hep-th/9802121,  submitted
to Class. Quant. Grav.

\bibitem{GR}
D.V. Gal'tsov and O.A. Rytchkov, Generating Branes via
Sigma models, hep-th/9801180.

\bibitem{IMS}
V.D.Ivashchuk and V.N.Melnikov,
Multidimensional Gravity with Einstein Internal spaces,
hep-th/9612054; {\it Gravitation and Cosmology}, 1996,
2, No. 3 (7),  211 (1996).

\bibitem{torino}
P. Fr\'e, Solvable Lie Algebras, BPS Black Holes and Supergravity
Gaugings, hep-th/9802045; \\
L. Castellani, A. Ceresole, R. D'Auria, S. Ferrara, P. Fr\'e
and M. Trigiante,$G/H$ M-branes and $AdS_{p+2}$ Geometries, hep-th/9803039.

\end{thebibliography}
\end{document}